# A Communication Avoiding and Reducing Algorithm for Symmetric Eigenproblem for Very Small Matrices

Takahiro Katagiri †, Jun'ichi Iwata ††, Kazuyuki Uchida ††

In this paper, a parallel symmetric eigensolver with very small matrices in massively parallel processing is considered. We define very small matrices that fit the sizes of caches per node in a supercomputer. We assume that the sizes also fit the exa-scale computing requirements of current production runs of an application. To minimize communication time, we added several communication avoiding and communication reducing algorithms based on Message Passing Interface (MPI) non-blocking implementations. A performance evaluation with up to full nodes of the FX10 system indicates that (1) the MPI non-blocking implementation is 3x as efficient as the baseline implementation, (2) the hybrid MPI execution is 1.9x faster than the pure MPI execution, (3) our proposed solver is 2.3x and 22x faster than a ScaLAPACK routine with optimized blocking size and cyclic-cyclic distribution, respectively.

## 1. Introduction

Currently, highly ranked supercomputers have more than 1,500,000 cores[1]. However, it is difficult to run large problems such as simulations per node, since the total problem size is huge due to the computational complexity. Therefore, we focus on a very small problem from the viewpoint of node capability.

Current numerical libraries are designed to attain high performance with huge matrix sizes. Level 3 operations, such as the dgemm routine in Basic Linear Algebra Subprograms (BLAS) are typical examples. These do not fit the design space for exa-scale computers. In exa-scale computers, the total number of nodes is on the order of 1,000,000. For a dense library, exa-scale computers require more than $N = 10,000$ sizes of matrices to obtain saturated performance per node. Since the matrix is two-dimensional (2D), the size of the matrix required in exa-scale computers reaches the order of 10,000 x $10{,}000 \times \sqrt{10{,}000{,}000}$ = 10,000,000. Since most dense solvers require $O(N^3)$ for computational complexity, the execution time with a matrix of $N = 10,000,000$ is unrealistic in actual applications. From the node viewpoint, the size of the target matrix should be limited to very small sizes. As a general estimation, we assume that $N = 1,000$ per node is the maximum size. The exa-scale size is on the order of $N = 1,000,000$. If this assumption is true, the used memory size of a matrix per node is only on the order of 8 MB, which is the cache size for current CPUs. In this study, we set a very small problem size, namely, the cache sizes for current CPUs.

### 1.1 RSDFT and Its Properties

Our target application is real space density functional theory (RSDFT) code[2]. RSDFT can perform quantum mechanical calculations of electronic structures with the first-principle method. The three-dimensional physical coordinates are discretized in RSDFT. Wave functions, electron densities, and potential fields are calculated at the resulting discrete grid points by solving a large-sparse eigenproblem of the dimension of the number of grid points. The algorithm of RSDFT is almost same as that of traditional plane-wave basis method; They are constructed from Conjugate Gradient (CG) minimization of Rayleigh quotients, Gram-Schmidt orthogonalization of wave functions, and a small-dense eigenproblem of the dimension of the number of wave functions necessary to calculate electron densities.

One of the advantages of RSDFT is that an FFT-free algorithm can be used. An FFT-free algorithm needs no global communications, which means not using collective communications with all Message Passing Interface (MPI) processes in distributed parallel execution. Moreover, because RSDFT uses a parallel algorithm with a 2D processor grid for the grid-point and wave function parallelization, simultaneous calls of collective communications are possible with a one-dimensional process grid.

Previous work[2] showed that RSDFT is scalable up to 442,368 cores (55,296 nodes) with 229,824 dimensions of a matrix in the K-computer. Due to their computational complexity, the eigensolver and orthogonalization parts become bottlenecks. The eigensolver and orthogonalization require $O(N^3)$ computations, while others require $O(N^2)$ computations. Hence, the eigensolver is one of the most important factors influencing speedup in massively parallel executions.

Parallel processing for RSDFT needs to consider whole process parallelism. As we mentioned, RSDFT needs to parallelize the orthogonalization routine and other routines, such as updating wave function using the CG method and updating potential fields, in addition to the eigensolver[2]. Although the number of processes may be too large for the eigensolver, it is necessary for the parallelization of other parts of RSDFT, and it is favorable to avoid any extra costs such as matrix re-distribution. Hence, matrices are distributed to obtain high parallelism for whole process execution. Since RSDFT has several target parts, the costs for the re-distribution of matrices is high. For this reason, we need distributed data input for the eigensolver, even though the target matrices are very small.

In addition, the eigensolver and other parts of RSDFT calculations are iterated more than hundreds of times until self-consistent potential fields are obtained. Hence, the target size of the matrices is very small with respect to the total execution time. Although the matrices are small and distributed, we need to speed up each part. Hence, the total execution time becomes a very severe situation for parallel processing.

In this paper, we discuss the adaptation of a new eigensolver to match the production sizes in RSDFT. In the case of the K-computer, the matrix size per node is very small at approximately 980x980. This is the same order that we mentioned in this section.

### 1.2 Originalities

The originalities of this paper are summarized as follows.
1. Communication avoiding and communication reducing algorithms are proposed for tridiagonalization and inverse transformation of symmetric eigensolvers.
2. The evaluation takes into account the problem size in a production run of a real application. For a conventional performance evaluation, a very small matrix is used.
3. Hybrid MPI execution performance for a symmetric eigensolver is evaluated with the full system of a peta-scale supercomputer consisting of 4,800 nodes (76,800 cores).

This paper is organized as follows. Section 2 explains the ABCLib_DRSSED[12] basics for implementations between communications and computations. Section 3 shows the performance evaluation results for a full-node system of the Fujitsu FX10 installed at the University of Tokyo. Section 4 summarizes the findings of this paper.

## 2. A Parallel Eigensolver: ABCLib_DRSSED

### 2.1 Related Work of Distributed Parallel Symmetric Eigensolver

Much effort has been directed at parallel implementation of a symmetric eigensolver. Historically, early parallel eigensolvers used a one-dimensional process grid. The communication was hidden by using previous elimination techniques[3] for the pivot vector (Householder vector) of tridiagonalization. Later, reduced communication algorithms using a 2D square grid were proposed[4)5]. Currently, we have two approaches for reducing the communication for tridiagonalization.

One is the one-step tridiagonalization approach, which is the same as that of conventional solvers. This approach uses a blocking algorithm for the main computations[6].

The two-step approach is the main design of the symmetric eigensolver. This approach uses a band reduction from a dense matrix to a banded matrix[7)8)9]. These approaches enable us to reduce the communication for the first step to the conventional one-step approach. Moreover, it uses a blocking algorithm to obtain high performance in the cache architecture for matrices of huge size. One of the drawbacks is the increased computation time for eigenvalue and eigenvector computations of the banded matrix. Another communication hiding approach for the two-step approach was proposed in reference 10).

As a result, all major algorithms use a blocking algorithm due to the design space for a huge matrix per node. As mentioned above, we assume that the design policy for exa-scale computers does not fit the current approaches. We use a totally different algorithm that uses a communication reducing algorithm based on the non-blocking algorithm[11].

**2.2 The Design Space**

The eigensolver provides a function that satisfies the following standard eigenvalue problem (SEP):

$$AX = X\Lambda, \qquad (1)$$

where $A \in \Re^{n \times n}$ is symmetric, $\Lambda \in \Re^{n \times n}$ is diagonal, and $X \in \Re^{n \times n}$ is orthogonal.

To solve SEP in (1), we employ a well-known numerical method that comprises the following three steps:

1. Tridiagonal reduction (TRD) step: Householder tridiagonalization, such as

    $A = QTQ^T$, where $Q = Q^T$,

2. Solving SEP with a Tridiagonal Matrix (SEPT) step: Solving SEP with a tridiagonal matrix $T$ by step (1), such as $T = V\Lambda V^T$, and

3. Householder inverse transformation (HIT) step: Householder inverse transformation with eigenvectors $V$ by step (2), such as $X = QV$.

In the design space of ABCLib_DRSSED, we employ the following design policy to reduce communications in the solver:

1. Use an implementation with reduced communications. This is adapted to an MPI non-blocking implementation in TRD.
2. Minimize the communication time in SEPT with a tridiagonal matrix.
3. Use reduced communication or an MPI non-blocking implementation in HIT.

Hereafter, we use the word ***communication*** to indicate executions with MPI functions.
In some situations, we also use ***communication*** to indicate processes for data packing and data unpacking in addition to executions of MPI functions.

Our target matrices are very small. Therefore, we need to manage the load imbalance for distributing matrix $A$. To avoid a poor load imbalance, we take the following approach:

- the data distribution of matrix $A$ is a cyclic-cyclic distribution with a blocking factor of 1.

We assume that all data per node can be put in cache memory space. Hence,

- a non-blocking algorithm is used to remove the copy time for data movement of the blocking algorithm.

For example, the copy time for data movement to use the BLAS3 routine generates overhead for some blocking algorithms. In our policy, because we do not use any BLAS routines, it is possible to implement BLAS-free computations in non-blocking algorithms. Hence, the copy time has no overhead.

Our target is minimizing the communication time when the ratio of computational time to that of communication is negligible.

## 2.3 Data Distribution

### 2.3.1 Matrix $A$

We use a 2D cyclic-cyclic distribution of matrix $A$. We map the MPI process onto a 2D grid of a $P_x \times P_y$ process.

The definition of the 2D cyclic-cyclic distribution is as follows.

Let the number of MPI processes be 2D, such as $P = P_x \times P_y$. The structure of $P_x \times P_y$ is called a process grid. The process identifier is also defined in 2D. That is, $(myid_x, myid_y)$ ranges from 0 to $P_x - 1$ for $myid_x$ and from 0 to $P_y - 1$ for $myid_y$.

The symmetric elements of dense matrix $A$ are distributed without symmetric compression. This means that all matrix data are stored and updated even though the matrix is symmetric. This supports the minimizing communication of matrix-vector and matrix update operations for TRD processes[5].

In the cyclic-cyclic distribution, indexes of the rows and columns for matrix $A$ are distributed as

$$\Pi = \{myid_x + 1 + (i-1)P_x\}, \quad \Gamma = \{myid_y + 1 + (j-1)P_y\}, \quad (2)$$

where

$$i = 1, 2, \cdots, last((myid_x + 1 + floor(n/P_x))P_x, floor(n/P_x)), \quad (3)$$
$$j = 1, 2, \cdots, last((myid_y + 1 + floor(n/P_y))P_y, floor(n/P_y)),$$

where

$$last(a, b) = \begin{cases} b+1 & (if\ a \leq n) \\ b & (if\ a > n) \end{cases}. \quad (4)$$

### 2.3.2 Matrices $T, V, X$ and $\Lambda$

The tridiagonal matrix $T$ is duplicated for all processes. The memory space of $T$ is $O(N)$ to $O(N^2/P_x)$ for $A$, since we accept the memory space for $T$.

Moreover, matrices $V$ and $X$ are not distributed for rows, but their columns are distributed. This is called a 1D distribution. The 1D distribution is also defined in the same way as formula (2), except $P$ is used instead of $P_x$. We set this 1D distribution of columns for $V$ and $X$ to reduce communications for computing $V$ in SEPT. For matrix $\Lambda$, the data distribution is the same as $V$ and $X$, since SEPT needs a corresponding matrix $\Lambda$ for $X$ for each process.

The 1D distribution of columns for $V$ and $X$ can reduce the communications for a process in SEPT, since if we use a 2D distribution, extra gathering or reducing operations for the distributed $V$ are needed as compared to the 1D distribution. If we use the parallel algorithm that requires no orthogonalization for each eigenvector of $T$, such as the MRRR algorithm, no communication occurs in SEPT. We introduce the MRRR algorithm in Section 2.7.

Please note that in the 1D distribution for $V$, $X$ with a 2D distribution for $A$ does not work for the ScaLAPACK routine, such as the PDSYEVD routine, since the specifications for the data distributions between $A$ and $X$ should be the same. Therefore, we think that our solver can reduce communications for SEPT to ScaLAPACK.

## 2.4 TRD Parallel Implementation

### 2.4.1 Notation of Vectors and Matrices

Let $A^{(k)}$ be a matrix $A$ in the $k$-th update.

Let $A_{i,j}$ be the element of matrix $A$ in the $i$-th row and $j$-th column. Let $A_{*,k}$ be a vector that is composed from the $k$-th column of $A$.

Let $A_{i:j,k:l}$ be a sub-matrix of $A$ that is composed from the $i$-th to $j$-th rows and $k$-th to $l$-th columns. The subscript of $A$ can be a set. For example, if $\Pi$ is a set of indexes of rows, $A_{\Pi,k}$ means the vector composed by the $k$-th row of $A$ with element locations specified by the elements of the set $\Pi$.

As in the matrix, the subscripts of vectors can be sets. For example, if $\Pi$ is a set of indexes of the elements of vectors, $x_\Pi$ means the elements of $x$ whose locations are specified by elements of the set $\Pi$.

The Householder matrix $Q$ can be defined as

$$Q = Q_{n-2} Q_{n-3} \cdots Q_1.$$

**2.4.2 Sequential Algorithm of TRD**

Given the notation in Section 2.4.1, the sequential algorithm of TRD can be described as follows.

$$A^{(k)}_{k:n,k} \to (\tau_k, v_k), \tau_k \in \Re, v_k \in \Re^{n-k+1}, \quad (5)$$

$$y_k^T = \tau_k v_k^T A^{(k)}_{k:n,k:n}, \quad (6)$$

$$\mu_k = \tau_k y_k^T v_k, \quad (7)$$

$$x_k = y_k, \quad (8)$$

$$Q_k A^{(k)} Q_k = A^{(k)} - (x_k - \tau_k v_k) v_k^T - v_k y_k^T, \quad (9)$$

where $k = 1, \cdots, n-2$. The map of $A^{(k)}_{k:n,k} \to (\tau_k, v_k)$ is the Householder reflection.

**2.4.3 Skeleton of TRD Process**

Figure 1 shows the kernel structure of the TRD process in ABCLib_DRSSED. As mentioned, we do not use the block algorithm of the Householder tridiagonalization. We also do not use the BLAS routine. Hence, the computations in ⟨8⟩-⟨10⟩ and ⟨18⟩-⟨22⟩ form loops in the computations. However, we use auto-tuning (AT) techniques [11] to optimize the codes, and the codes are tuned by an auto-tuner in ABCLib_DRSSED.

In Figure 1, we portray the send operation as **Send**, and the receive operation as **Receive**.

```
⟨ 1⟩ do k = 1, n − 2
⟨ 2⟩   if (k ∈ Γ) Send (A^(k)_Π,k) to processes sharing rows Π.
⟨ 3⟩   else Receive (A^(k)_Π,k) endif
⟨ 4⟩   Computation of (τ_k, (v_k)_Π) with MPI_Allreduce.
⟨ 5⟩   if (I have diagonal elements of A ) then
⟨ 6⟩     Send ((v_k)_Π) to processes sharing columns Γ.
⟨ 7⟩   else Receive ((v_k)_Γ) endif
⟨ 8⟩   do j = k, n
⟨ 9⟩     if (j ∈ Π) ((y_k)_Π)^T = ((y_k)_Π)^T + τ_k (v_k^T)_j A^(k)_Π,j
⟨10⟩   enddo
⟨11⟩   MPI_Allreduce of ((y_k)_Π)^T to processes sharing rows Π.
⟨12⟩   if (I have diagonal elements of A ) then
⟨13⟩     Send ((y_k)_Π) to processes sharing columns Γ. (x_k)_Γ = (y_k)_Π.
⟨14⟩   else Receive ((x_k)_Γ).
⟨15⟩   do j = k, n
⟨16⟩     μ_k = τ_k ((y_k)_Π)^T (v_k)_Π;  enddo
⟨17⟩   MPI_Allreduce of μ_k to processes sharing rows Π.
⟨18⟩   do j = k, n
⟨19⟩     do i = k, n
⟨20⟩       if (i ∈ Π .and. j ∈ Γ) then
⟨21⟩         A^(k+1)_i,j = A^(k)_i,j − (v_k)_i ((x_k^T)_j − μ_k (v_k^T)_j) − (v_k)_i (y_k^T)_j; endif
⟨22⟩   enddo; enddo
⟨23⟩   if (k ∈ Γ) Γ = Γ − { k }; endif
⟨24⟩   if (k ∈ Π) Π = Π − { k }; endif
⟨25⟩ enddo
```

Fig. 1 Skeleton of TRD process in ABCLib_DRSSED.

These send and receive operations have several implementations, e.g., MPI_Bcast. The performance of these operations depends on the configuration of the processor grids.

Computation $(\tau_k, (v_k)_\Pi)$ in ⟨4⟩ is called the Householder reflection, and the vector is called the pivot vector of the Householder reflection. The pivot vector is sent to or received by processes that share rows by the communication in ⟨2⟩-⟨3⟩ to reduce communication times.

Alternatives are possible for communication implementations in ⟨5⟩-⟨7⟩ and ⟨12⟩-⟨14⟩ according to the process grids. If the process grid is square, these communications are optimized to use multi-casting MPI_Bcast. This algorithm is known as the scattered square algorithm[4)5)].

In actual use, it is difficult to use a square grid. Therefore, the process grid should be rectangular. Our implementation changes communication in ⟨5⟩-⟨7⟩ and ⟨12⟩-⟨14⟩ according to the organization of the process grid. To reduce the communication time, we use multiple MPI_Allreduce()s rather than MPI_GatherV(). (See reference 10) for details of the implementations.)

As mentioned above, the configuration of the process grid affects the communication time.

Hence, the tuning of process grids is a very important issue for performance tuning in the TRD process.

The above two algorithms are communication avoiding algorithms, since communications for the pivot vector (Householder vector $v_k$) can be avoided in the original algorithm by the redundant $v_k$ values for each process sharing column $\Gamma$.

## 2.5 A New Implementation with MPI Non-blocking Communication in HIT

A communication overlapping approach with MPI non-blocking communication in TRD is proposed in this paper. Details of the approach are shown in Figure 2. The aim of our approach is to reduce synchronization costs by replacing MPI synchronization (blocking) communications with MPI non-blocking communications.

```
    if (k .le. K_PrevSend) then
⟨ 2'⟩   if (k ∉ Γ) Receive (A^(k)_{Π,k}); endif
...
⟨18'⟩  if (k .le. K_PrevSend) then
⟨19'⟩    if (I have elements of A^(k+1)_{*,k+1}) then
⟨20'⟩      do j = k, k + 1
⟨21'⟩        do i = k, n
⟨22'⟩          if (i ∈ Π .and. j ∈ Γ) then
⟨23'⟩            A^(k+1)_{i,j} = A^(k)_{i,j} - (v_k)_i ((x_k^T)_j - μ (v_k^T)_j) - (v_k)_i (y_k^T)_j; endif
⟨24'⟩        enddo; enddo
⟨25'⟩     call MPI_Isend (A^(k+1)_{Π,(k+1)}) to processes sharing rows Π.
⟨26'⟩     do j = k + 2, n
⟨27'⟩       do i = k, n
⟨28'⟩         if (i ∈ Π .and. j ∈ Γ) then
⟨29'⟩           A^(k+1)_{i,j} = A^(k)_{i,j} - (v_k)_i ((x_k^T)_j - μ (v_k^T)_j) - (v_k)_i (y_k^T)_j; endif
⟨30'⟩       enddo; enddo
⟨31'⟩  endif
```

Fig. 2 MPI Non-blocking implementation of sending pivot vector. The lines denoted by prime symbols replace the corresponding lines without the prime symbols in Figure 1.

In Figure 2, the pivot vector of the next iteration step is calculated be forehand in lines ⟨ 20' ⟩ - ⟨ 24' ⟩. Then, the pivot vector is immediately sent with MPI_Isend() in lines ⟨ 25' ⟩. The aim of this implementation is to hide the sending time for the pivot vector. This is established by the computation in lines ⟨ 27' ⟩ - ⟨ 31' ⟩.

The above sending is effective if the iteration is less than or equal to K_PrevSend. The parameter K_PrevSend, hence, is a tunable parameter in TRD.

## 2.6 HIT Parallel Implementation
### 2.6.1 Sequential Algorithm of HIT

The sequential algorithm of HIT can be described as follows.

$$\sigma_i = \tau_k v_k^T X_{k:n,i}^{(k)}, \quad (10)$$

$$Q_k X_{k:n,i}^{(k)} = X_{k:n,i}^{(k)} - \sigma_i v_k, \quad (11)$$

where $i = k, \cdots, n$, and $k = n-2, \cdots, 1$.

### 2.6.2 Skeleton of HIT Process

Figure 3 shows the kernel structure of the HIT process in ABCLib_DRSSED. The HIT kernel is blocked for gathering the distributed pivot vectors for $v_k$. The data distribution of the pivot vectors is the cyclic-cyclic distribution across $P_x$ processes. Hence, the vectors are copied for processes that share columns $\Gamma$. The data source for sending pivot vectors is reduced by referring to the copied (redundant) pivot vectors. Hence, the algorithm in Figure 3 is a communication avoiding algorithm that utilizes redundant pivot vectors.

```
⟨ 1⟩ do k = n − 2, 1, −MBLK
⟨ 2⟩   do kbl = k, max(k − MBLK + 1,1), −1
⟨ 3⟩     Allgather the vector v_{kbl} with sharing columns Γ.
⟨ 4⟩   enddo
⟨ 5⟩   do kbl = k, max(k − MBLK + 1,1), −1
⟨ 6⟩     do i = kstart, kend
⟨ 7⟩       σ_i = τ_k v_k^T X^(kbl)_{kbl:n,i}
⟨ 8⟩       X^(kbl+1)_{kbl:n,i} = X^(kbl)_{kbl:n,i} − σ_i v_{kbl}
⟨ 9⟩     enddo
⟨10⟩   enddo
⟨11⟩ enddo
```

Fig. 3 Skeleton of HIT process in ABCLib_DRSSED. The global index for the first location is *kstart* and the global index for the last location for $v_k$ is *kend*.

For implementation of the gathering part in ⟨ 3 ⟩, several implementations can be taken into account. A conventional implementation for ABCLib_DRSSED uses MPI_Bcast() with $P_y$-times calling. This implementation is shown in Figure 4. In Figure 4, "MPI_Bcast() with $P_y$-times calling" is a way to implement an Allgather.

⟨3′⟩ **Allgather** the vector $v_{kbl}$ by using $P_x$-times of MPI_Bcast for $(v_{kbl})_\Pi$ with sharing columns $\Gamma$.

Fig. 4  MPI Bcast implementation in the HIT process.

Configuration of the process grid affects the communication time in the HIT process, since the number of simultaneous MPI_Bcasts depends on $P_x$. Hence, tuning of the process grids is also a very important issue for performance tuning in the HIT process.

**2.6.3 A New Implementation with MPI Non-blocking Communication and Alternative Implementation in HIT**

In this paper, we propose two kinds of implementations for the gathering process in HIT. The first implementation is an MPI non-blocking communication version, explained in Figure 5.

In Figure 5, sending the pivot vector is performed asynchronously in ⟨3′-1⟩. Hence, the sending process overlaps the computation in ⟨4⟩-⟨6⟩ shown in Figure 3. The blocking factor, MBLK, can also be a tunable parameter for the MPI non-blocking communication.

⟨3′-1⟩ **if** (I have $(v_{kbl})_\Pi$) **then**
⟨3′-2⟩   **Send** the vector $v_{kbl}$ by using $P_x$-times of MPI_Isend for $(v_{kbl})_\Pi$ with sharing columns $\Gamma$.
⟨3′-3⟩ **else**
⟨3′-4⟩   call MPI_Recv(···); **endif**
...
⟨3′-5⟩ **if** (I have $(v_{kbl})_\Pi$) call MPI_Wait(···);

Fig. 5  MPI non-blocking implementation of the gathering process in HIT.

Next, we implement the alternative gathering process with MPI_BCAST(). Figure 6 shows this implementation.

⟨3′-1⟩ **do** $ii = 1, P_x$
⟨3′-2⟩   **do** $kbl = k, \max(k - MBLK + 1, 1), -1$
⟨3′-3⟩     Copy $(v_{kbl})_\Pi$ to a sending buffer $VS_{*,kbl}$.
⟨3′-4⟩   **enddo**
⟨3′-5⟩   **Allgather** the vectors $v_k, \cdots, v_{\max(k-MBLK+1,1)}$ by using MPI_Bcast for $VS_{*,k:\max(k-MBLK+1,1)}$ with sharing columns $\Gamma$.
⟨3′-6⟩ **enddo**

Fig. 6  Alternative communication implementation of the gathering process in HIT (block MPI_Bcast).

In Figure 6, the loops between the MPI blocking and gathering processes that are related in the $P_x$ processes are exchanged. The implementation of Figure 6 is effective if the number $P_x$ is less than the value of the blocking factor MBLK.

Figure 6 is a communication reducing algorithm, since the number of communications is reduced by a factor of 1/MBLK compared to that of the original algorithm.

The original algorithm in Figure 3 is a communication avoiding algorithm, since the communications for the pivot vector (Householder vector $v_k$) can avoid having a duplicate copy of $v_k$ for each process. This is not a normal parallel implementation, since if we have a duplicate copy of vk for each process, we need extra-large memory spaces. The extra memory space is $O(N^2/P_x)$ in the $P_x \times P_y$ processor grid. This means that if we use the 1D distribution with $1 \times P$ processor grids, all elements of the Householder vectors are duplicated for all MPI processes. This also implies that the 1D distribution should have very limited problem sizes. Hence, to reduce memory space, Householder vectors are not redundant in a normal parallel implementation. If Householder vectors are not redundant, we need some communications. Hence, we call this a communication avoidance algorithm.

The algorithm in Figure 3 was proposed in our previous work[11]. In this paper, we propose a mix between communication avoidance for the algorithm in Figure 3 and communication reducing in Figure 6.

Figure 7 shows a snapshot of the algorithm in Figure 6 when the size of communication blocking is 2 in a 2x4 process grid.

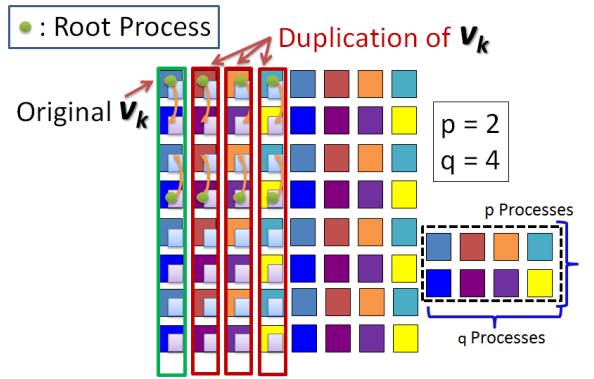

(a) Conventional communication for $v_k$.

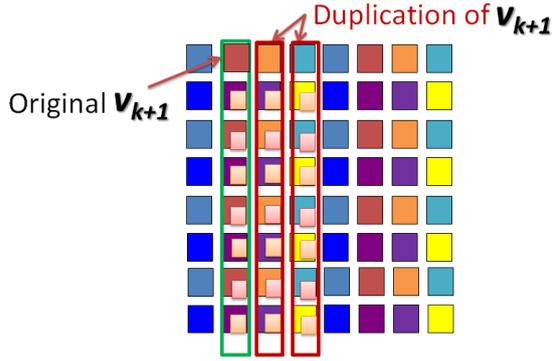

(b) Data distribution for next Householder vector of $v_{k+1}$.

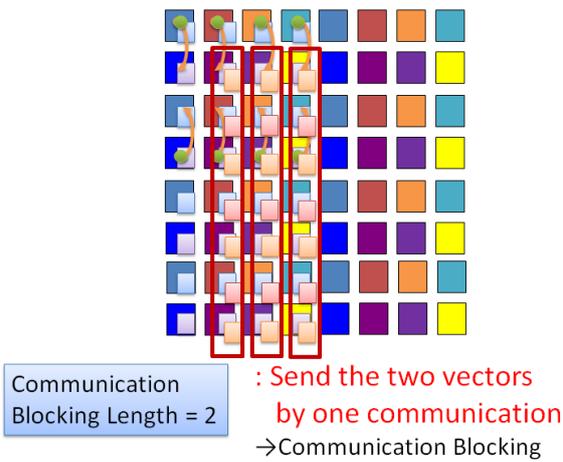

(c) Communication blocking for vectors $v_k$ and $v_{k+1}$.

Fig. 7  A snapshot of communication blocking for HIT. Process grid is 24. Size of communication blocking is 2.

**2.7 Thread Parallelization for SEPT**

To minimize the communication in SEPT, we use the sequential Multiple Relatively Robust Representatives (MRRR) algorithm[13] for each process that is called by the parallel implementation.

In a conventional MRRR implementation, the bisection method is used to compute eigenvalues. For this reason, it is difficult to accelerate the computation in thread execution, since the bisection method does not support parallelism. Hence, MRRR causes a problem for hybrid MPI execution.

To solve this problem, we use the Multi-Eigenvalues and Multi-Section (MEMS) implementation for MRRR[14]. MEMS uses a multi-section method for computing one eigenvalue, and it can also be used for simultaneous computations of eigenvalues. MEMS has two tunable parameters in the MRRR routine: (1) the number of multi-sections (ML), and (2) the number of simultaneous computations of eigenvalues (EL).

## 3. Performance Evaluation

### 3.1 Machine Environment

#### 3.1.1 Hardware Overview

We use the Fujitsu FX10 installed in the Information Technology Center at The University of Tokyo. The total number of nodes of the FX10 is 4,800, and the total number of FLOPS is 1.135 PFLOPS. The node architecture is the SPARC64 IX-fx (1.848 GHz) that provides 235.5 GFLOPS per node, 16 cores, and 32 GB memory. Hence, the total number of cores in the full system of the FX10 is 76,800. The SPARC64 IX-fx has two hierarchical caches, a separated L1 cache and a shared L2 cache. The sizes of the L1 and L2 caches are 32 KB and 12 MB, respectively.

The inter-connect network is the TOFU network that Fujitsu developed for a dedicated six-dimensional mesh. The link throughput of the TOFU network is 5 GB/sec in each direction. We use full nodes of the FX10 for this evaluation.

#### 3.1.2 Software Overview

For the software environment, we use the Fujitsu Fortran90 compiler version 1.2.1 with the option "-Kfast, -openmp". The Fujitsu-optimized ScaLAPACK (version 1.8) and the Fujitsu-optimized BLAS are also used for the experiments in the ScaLAPACK routine. These libraries are released as the Fujitsu SSLII library.

For the MRRR routine in parallel execution, we use an equivalent implementation for the LAPACK version 3.0 xLARRD implementation, but we modified it to add MEMS from the original bisection routine in MRRR.

Please note that the accuracy of the over-process eigenvalue and eigenvector computations is not guaranteed in the sequential execution in our implementation, since our implementation, MRRR with MEMS, is called with only our own inner eigenvalues and eigenvectors. The accuracy of the eigenvectors is retained if the relative distance of the eigenvectors is accepted for the condition of the MRRR algorithm.

We evaluate pure MPI and hybrid MPI executions. For pure MPI execution, we use MPI processes to match the numbers of all cores. For hybrid MPI execution, we have several choices for execution patterns. In this experiment, we fix the threads for the maximum number of threads per node. That is, we use 16 threads per node in the FX10. The reason for this limitation is to reduce the time for inter-node communications. This is the most important goal for hybrid MPI execution.

## 3.2 Test Matrix

### 3.2.1 Target Matrix

The target process of this evaluation is to calculate all eigenvalues and all eigenvectors with a given matrix. The matrix is limited to a square symmetric real matrix.

We used the Frank matrix to check computed eigenvalues and eigenvectors. The Frank matrix is defined as follows.

$$A = (a_{ij}), a_{ij} = n - \max(i,j) + 1. \quad (12)$$

The analytical eigenvalues are as follows.

$$\lambda_k = \frac{1}{2(1 - \cos\frac{2k-1}{2n+1}\pi)}, (k = 1, 2, \cdots, n). \quad (13)$$

We also check for errors in the calculated eigenvalues with formula (13).

### 3.2.2 Time Measurement Points

We measured the time for the following processes. All time is summed at each outer loop iteration until the loops are finished.

- TRD routine
  - Communication Time
    1. **Send Piv**: 〈2〉-〈3〉 in Figure 1. Time of data packing and copy back is included.
    2. **Send yt**: 〈5〉-〈7〉 in Figure 1. Time of data packing and copy back is included.
    3. **Send xt**: 〈12〉-〈13〉 in Figure 1. Time of data packing and copy back is included.
    4. **MatVec Reduce**: 〈11〉 in Figure 1.
  - Computation Time
    1. **Matvec**: 〈8〉-〈10〉 in Figure 1.
    2. **Update**: 〈18〉-〈22〉 in Figure 1.
    3. **Other**: Other time for the above processes.
- HIT routine
  - Communication Time
    1. **Send Piv**: 〈2〉-〈4〉 in Figure 3. Time for data packing and copy back is included.
  - Computation Time
    1. **HIT Ker**: 〈5〉-〈10〉 in Figure 3.
    2. **Other**: Other time for the above processes.

The time breakdown shown is for the last process in the process identifier. If 1,024 processes are used, the time breakdown of the 1,023 processes is presented.

### 3.2.3 Target Matrix Sizes

As we mentioned, the target size of the matrix is cached sizes for the shared L2 cache (12 MB) in the FX10. Due to this small size, we fix $N$ = 600 per node. In this situation, we need 2.74 MB for matrix $A$. We can increase the maximum size up to $N$ = 1,200 per node, since that case requires 10.9 MB per node.

The matrix is 2D. Hence, we increase the number of nodes by a factor of 2 according to the matrix size. As a result, we fix the target sizes as follows: $N$ = 1,200 (4 nodes), $N$ = 2,400 (16 nodes), $N$ = 4,800 (64 Nodes), $N$ = 9,600 (256 nodes), $N$ = 19,200 (1,024 nodes), and so on.

## 3.3 AT Function and Parameter Searching

ABCLib_DRSSED[12] has an AT function for computations and communications. In this experiment, the AT function of the computations can be ignored, since all data are in the cache. Here, we explain the AT function for communications.

We implemented the following AT candidates in the experiment.
- TRD routine
  - #1. Binary-tree implementation with MPI Send and MPI_Recv in < 11 > in Figure 1.
  - #2. MPI_Bcast implementation in < 11 > in Figure 1.
  - #3. MPI Blocking implementation in < 2 > - < 3 > in Figure 1 or MPI non-blocking implementation in Figure 2.
- HIT routine
  - #1. MPI_Bcast implementation in Figure 4.
  - #2. MPI non-blocking implementation in Figure 5.
  - #3. Alternative MPI_Bcast implementation in Figure 6.

In this experiment, the parameter estimation function is not used: all implementations are searched by the input matrix size. The blocking factor *MBLK* is a tunable parameter in HIT. We limit the search space of *MBLK* as follows.
- *MBLK* ∈ {1, 2, 4, 8, 12, 16, 32, 48, 56, 64, 80, 96, 112, 128}

Due to the blocking factor, it is difficult to search all combinations. The total combinations of the parameter space are 16 x 3 = 48 points for one dimension. Hence, we use the following ad hoc searching method for HIT tuning.
1. Set the communication implementation to #1 (MPI_Bcast in Figure 4).
2. Search the best block factor.
3. Search the best implementation between #1 (MPI_Bcast), #2 (MPI non-blocking implementation), and #3 (Alternative MPI_Bcast implementation) with the optimized blocking factor in step 2.

### 3.4 TRD Performance in 64 Nodes (1,024 Cores)
#### 3.4.1 Pure MPI Execution
Figure 8 shows the TRD execution time for pure MPI execution in various process configurations (e.g., 2 x 512).

According to Figure 8, the case of the square grid (32 x 32) dramatically reduces the communication time for `send xt`. This is due to the optimal MPI_Bcast, which supplies the most efficient communication for a square grid. The times of `Other` are increased for the other configurations. The times of `Other` are calculated by (global synchronous ending time) – (summation of all breakdowns).

One reason is that the time of waiting for each communication is accumulated for the time of `Other`. In the remaining cases, the time for the dot-product, including MPI_Allreduce, and the time for storing vector vk increase in the square grid. More analysis is needed to clarify this phenomenon. The 16 x 64 configuration is the fastest in this execution.

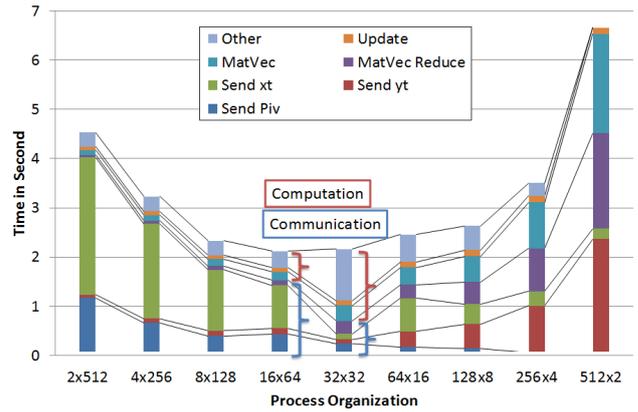

Fig. 8   TRD execution time in pure MPI execution in 1,024 processes (64 nodes).

#### 3.4.2 Hybrid MPI Execution
Figure 9 shows the TRD execution time for hybrid MPI execution in various process configurations.

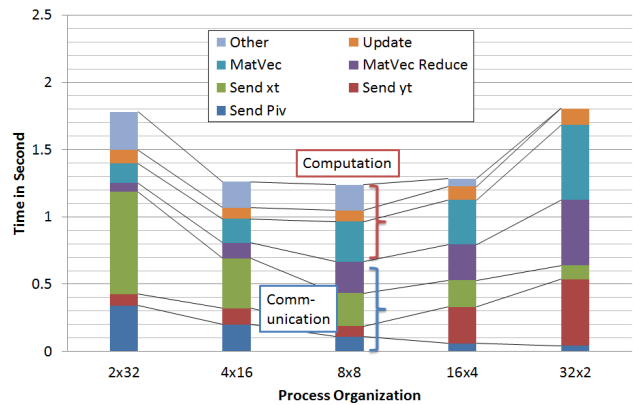

Fig. 9  TRD execution time in hybrid MPI execution

in 64 processes (64 nodes).

Figure 9 indicates that the square grid minimizes the execution time. The ratio between communication and computation to the total time is almost half. The heaviest process is `Matvec`.

### 3.5  HIT Performance in 64 Nodes (1,024 Cores)
#### 3.5.1 Pure MPI Execution

Figure 10 shows the HIT execution time for pure MPI execution in various process configurations.

Figure 10 indicates that the time for the send process is increased according to the number of processes and the increase of $P_x$. This is due to the time for gathering the distributed $v_k$. Hence, for the HIT routine, it is better to have a small number of $P_x$.

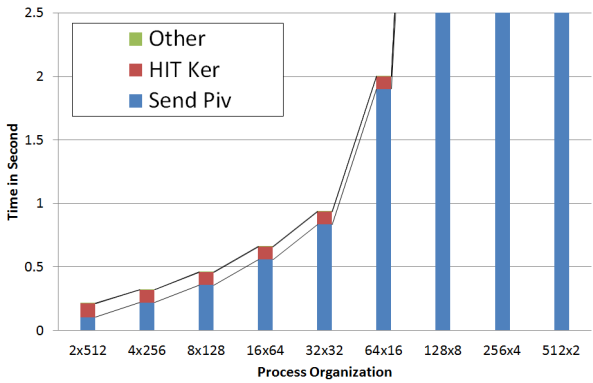

Fig.10  HIT execution time in pure MPI execution

in 1,024 processes (64 nodes).

#### 3.5.2 Hybrid MPI Execution

Figure 11 shows the HIT execution time for hybrid MPI execution in various process configurations.

The tendency of the execution time in Figure 11 is the same as that in Figure 10; however, the total time is shortened. This is the effect of hybrid MPI execution by reducing the total number of processes.

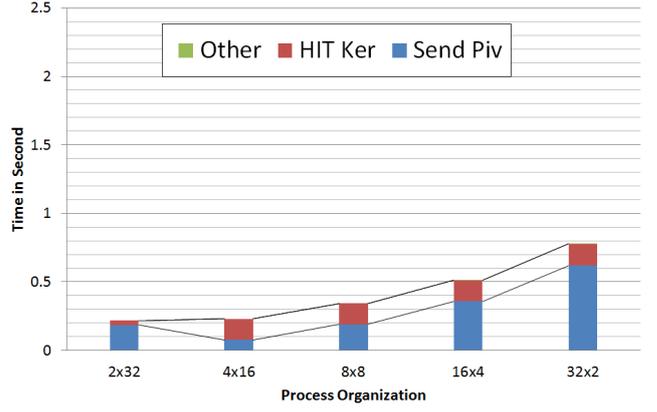

Fig.11  HIT execution time in hybrid MPI execution

in 64 processes (64 nodes).

### 3.6  Total Executions in 64 Nodes (1,024 Cores)
#### 3.6.1 Pure MPI Execution

Figure 12 shows the total execution time for pure MPI execution in various process organizations.

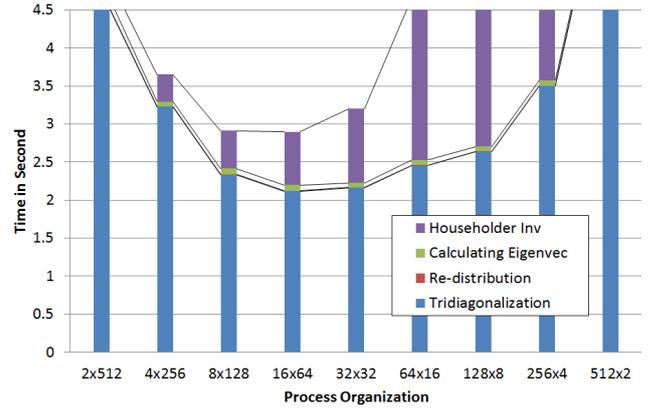

Fig.12  Total execution time in pure MPI execution

in 1,024 processes (64 nodes).

Figure 12 shows that the fastest grid is 16x64. One reason is that our implementation limits the increase of communications in the HIT routine if a small number of $P_x$ processes have redundant vectors of $v_k$ in each process.

#### 3.6.2 Hybrid MPI Execution

Figure 13 shows the total execution time for hybrid MPI execution in various process configurations.

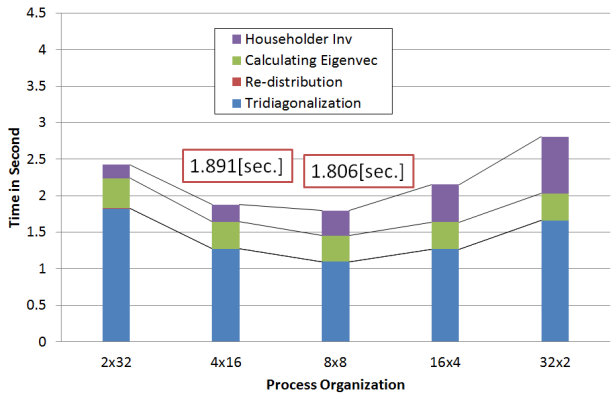

Fig.13　Total execution time in hybrid MPI execution in 64 processes (64 nodes).

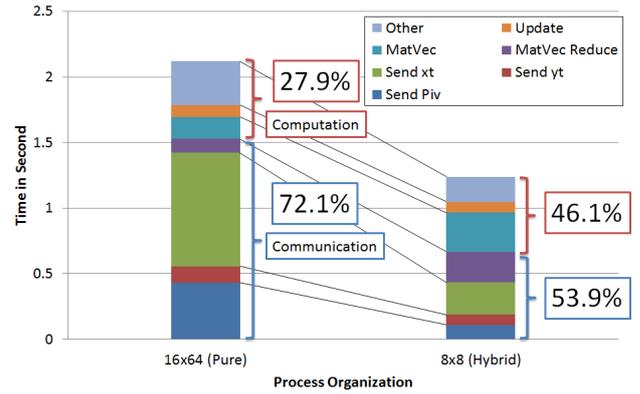

Fig.14　Comparison of the best pure and hybrid MPI executions of TRD in 64 nodes.

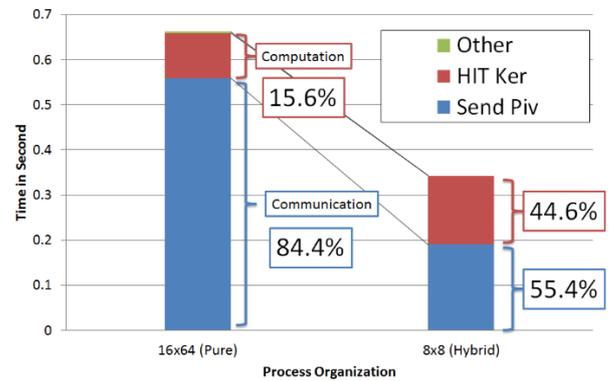

Fig.15　Comparison of the best pure and hybrid MPI executions of HIT in 64 nodes.

In Figure 13, the fastest process grid is 8x8. It is remarkable that the ratio of the calculation time for eigenvectors is higher than that of the pure MPI execution in Figure 12. This is because the parallelization of computing eigenvectors depends on the number of MPI processes. The parallelism is not enough in the hybrid MPI, although we use thread parallelism with MEMS for the MRRR routine.

**3.6.3** Comparison of Pure and Hybrid for the Best Execution Case

Figure 14 shows a comparison of the best times of pure and hybrid MPI executions in TRD.

Figure 14 indicates that the ratio of communication can be reduced from 72.1% to 53.9% by using hybrid MPI execution. The execution time can also be reduced from 2.11 [sec] to 1.24 [sec]. This implies that a 1.7x speedup of the pure MPI execution is established.

Figure 15 is a comparison of the best times between pure and hybrid MPI executions in HIT.

Figure 15 shows a significant reduction from 84.4% to 55.4% for the ratio of communication time obtained by using hybrid MPI. The time for reduction decreases from 0.66 [sec] to 0.34 [sec].

**3.7　Effect of Communication Tuning in 64 Nodes (1,024 Cores)**

**3.7.1** Communications in TRD

Figure 16 shows the effect of the MPI non-blocking communication and the previous elimination of pivot vectors in TRD.

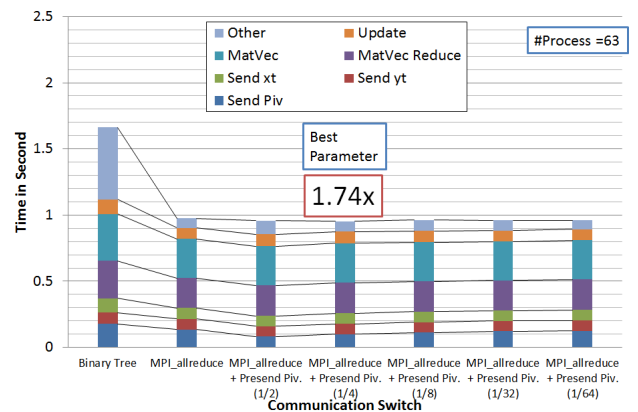

Fig.16　Effect of different communication implementations of TRD in hybrid MPI executions in 64 nodes.

Figure 16 shows that the previous elimination of pivot vectors has almost no effect on the MPI non-blocking communication,

whereas the MPI_Allreduce implementation is very efficient in the FX10.

This reduces the waiting time of binary tree communications, since the **Other** time includes the waiting time for each process. This result depends on the total number of processes and the matrix sizes. The best implementation uses an MPI non-blocking communication from 1/4 iterations to all iterations, which is 4,800/4 = 1,200 in this matrix. The times with MPI_Allreduce and MPI non-blocking with the 1/4 iterations are 0.975 [sec] and 0.953 [sec], respectively. Hence, the effect of MPI non-blocking is only a speedup of 2.3%.

### 3.7.2 Communications in HIT

Figure 17 shows the effect of the implementation candidates for communications in HIT.

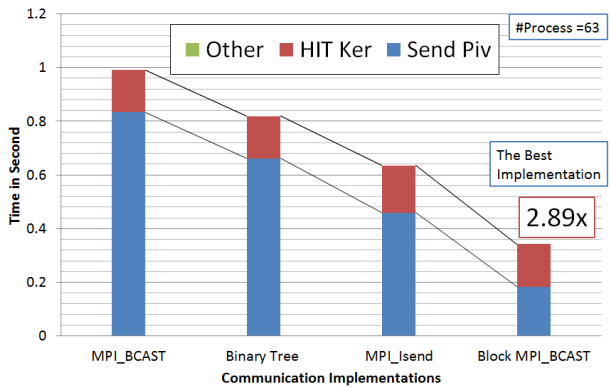

Fig.17 Effect of different communication implementations of HIT in hybrid MPI execution in 64 nodes.

Figure 17 shows the most effective implementation is a block implementation with MPI Bcast. The speedup factor over the simple MPI_Bcast implementation is 2.89x. The best implementation, of course, depends on the number of processes and matrix sizes. If the number of processes is increased to more than 64 processes, implementation of MPI_Isend is more effective, since the block MPI_Bcast needs synchronization with $O(P_x)$. (See the result in Section 3.9.)

Figure 18 shows the effect of the blocking factor for HIT. The implementation of the communication is fixed for block MPI_Bcast in Figure 18.

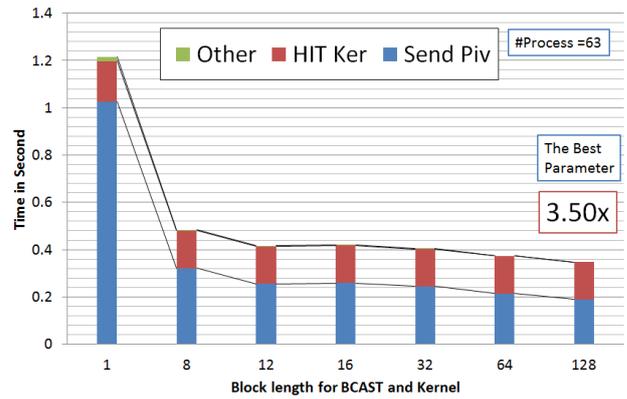

Fig.18 Effect of varying blocking factors in HIT in hybrid MPI execution in 64 nodes by using block MPI_Bcast.

According to Figure 18, blocking implementation is very crucial to the total performance in HIT. According to the MPI non-blocking implementation (blocking factor = 1) in Figure 18, the execution time is dramatically reduced to 3.5x with blocking factor = 128.

### 3.8 Tuning of MRRR Routine with MEMS

As we mentioned, we use thread parallelism for the MRRR routine[13] by using the MEMS method[14] and replacing the original bisection method in MRRR. In this experiment, we tuned the *ML* and *EL* parameters in MEMS. *ML* is the point of multi-section for eigenvalue counting, and *EL* is the number of eigenvalues for computing simultaneously. We find that the best parameters are *ML* = 2 and *EL* = 75 in this matrix. According to the original bisection, which implies execution with *ML* = 1 and *EL* = 1 in MEMS, we obtain a 1.16x speedup with 16 threads. Hence, we use the parameters *ML* = 2 and *EL* = 75 in this experiment.

### 3.9 Performance on 4,800 Node Execution (76,800 Cores)

Figure 19 shows full-node execution in the FX10 for hybrid MPI execution in TRD. Due to the limitation of experimental time for full-node execution in the FX10, we can only check for process configurations 20 x 240 and 40 x 120.

Figure 19 shows that the processor grid of 40 x 120 is faster than the 20x240 processor grid. The most reduced time is the time for **send piv**. This is because **send piv** depends on the number of $P_y$.

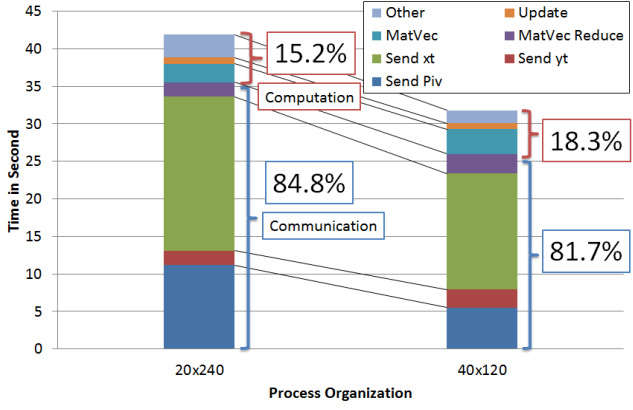

Fig.19 TRD execution time in hybrid MPI execution
in 4,800 processes (4,800 nodes, 76,800 cores).

Figure 20 shows the breakdown of full-node execution in the FX10 for hybrid MPI in the HIT. The blocking factor of 4 with MPI non-blocking communication is used for this result. This factor comes from the result of the AT function.

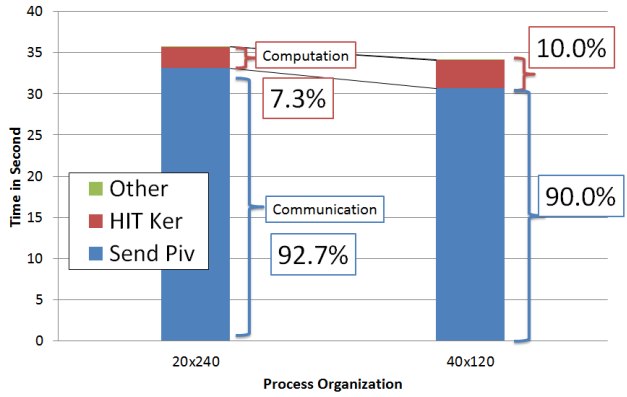

Fig.20 HIT execution time in hybrid MPI execution
in 4,800 processes (4,800 nodes, 76,800 cores).

Figure 20 indicates that process organization is not sensitive in HIT in comparison to that of TRD. The best process configuration is also 40 x 120.

Finally, we evaluate the performance under varying matrix sizes. The process configuration is fixed as 40 x 120. The result is shown in Figure 21.

Figure 21 indicates the following very important features. (1) The doubled dimension requires only a 3.97x increment of execution time for $N$ = 83,138, although the computational complexity of the algorithm is $O(N^3)$. (2) However, it needs 5.0x execution time in the case of $N$ = 166,276. The target of this paper is less than $N$ = 83,138, since the case of $N$ = 166,276 requires much memory space for the L2 cache in the FX10. As a result, we conclude that our solver is very effective for the size of the L2 cache.

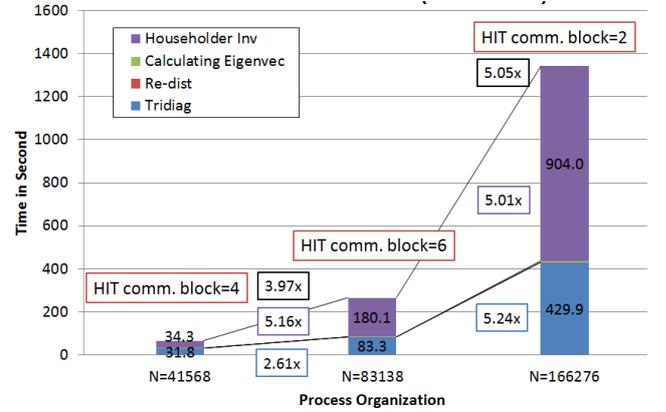

Fig. 21 Total execution time with varied matrix sizes in hybrid MPI execution in 4,800 processes (4,800 nodes, 76,800 cores).

In contrast, the ratio of execution time in HIT increases in comparison with the case of 64 nodes. For 4,800 nodes, the ratio between the execution time of TRD and HIT is almost the same as in the case of $N$ = 41,568. This means that the cost of gathering distributed vectors $v_k$ is a heavy task in massively parallel execution.

### 3.10 Comparison to ScaLAPACK Routine

We compare the best execution times between ABCLib_DRSSED and ScaLAPACK PDSYEVD routines. Pure MPI execution is evaluated for ABCLib_DRSSED and ScaLAPACK PDSYEVD in this comparison. Hence, BLAS in ScaLAPACK is sequential BLAS, but the BLAS is well optimized by the vender.

The PDSYEVD routine uses a divide-and-conquer algorithm for SEPT, while we use MRRR. This is because we do not have the vender-optimized ScaLAPACK version supplying the MRRR routine (PxSYEVR), which is supported in higher versions than 2.0. Table 1 shows the result.

According to Table 1, ABCLib_DRSSED speedups are from 1.65x to 2.37x for the best case of the ScaLAPACK routine. However, the data distribution is fixed as a cyclic-cyclic distribution in the application requirement, so this case should be fixed as *MBSIZE* = 1. ABCLib_DRSSED establishes speedup from 10.97x to 22.08x in the case of *MBSIZE* = 1. If an application requires cyclic-cyclic distribution, ABCLib_DRSSED is a very powerful tool for ScaLAPACKa.

Table. 1 Execution times between ScaLAPACK PDSYEVD and ABCLib_DRSSED. Time is in seconds. Pure MPI execution is used for ScaLAPACK PDSYEVD and ABCLib_DRSSED.

| N, Nodes (Process Grid in ScaLAPACK) | $N$ = 4800 64 (8x8) | $N$ = 9600 256 (16x16) | $N$ = 19200 1024 (32x32) |
|---|---|---|---|
| *MBSIZE* = 1 | 39.69 | 83.15 | 170.33 |
| *MBSIZE* = 8 | 6.84 | 15.41 | 35.28 |
| *MBSIZE* = 16 | 5.12 | 12.61 | 28.86 |
| *MBSIZE* = 32 | 4.58 | 11.16 | 26.96 |
| *MBSIZE* = 64 | 4.26 | 82.28 | 25.76 |
| *MBSIZE* = 128 | 4.39 | 10.96 | 26.35 |
| *MBSIZE* = 256 | 4.72 | 11.57 | 27.62 |
| ABCLib_DRSSED (Process Grid) | 1.79 (8x8) | 4.61 (8x32) | 15.52 (16x64) |
| Speedup to ScaLAPACK (The Best) | 2.37x | 2.37x | 1.65x |
| Speedup to ScaLAPACK (*MBSIZE* = 1) | 22.08x | 18.02x | 10.97x |

### 3.11 Accuracy

The accuracy of the computed eigenvalues and eigenvectors needs to be taken into account, since we change the dependency of the MRRR part of the sequential execution. First, we compared the maximum error of eigenvalues between the PDSYEVD routine and that of our solver in the case of $N$ = 19,200 with 1024 nodes. The result was 4.163E-07 (PDSYEVD with *MBSIZE* = 1) and 3.939E-10 (our proposed solver). Hence, our solver can establish better accuracy for the ScaLAPACK routine

For the maximum error of the orthogonality of eigenvectors and the maximum residual error of the eigensystem in the case of $N$ = 19,200 with 1024 nodes, we obtain 8.882E-10 (maximum error in Frobenius norm of $\|X^T X - I\|_F$) and 1.591E-08 (maximum error in 2-norm of $\|Ax_i - \lambda_i x_i\|_2, (i = 1,...,n)$, respectively. Although our algorithm does not follow the sequential data dependencies of the MRRR algorithm, the result is not a problem for the Frank matrix. One reason is our algorithm keeps the relative distance of eigenvalues between our eigenvalues for inter-processors in the Frank matrix.

### 4. Conclusion

In this paper, we evaluated a parallel symmetric eigensolver for very small matrices in massively parallel processing. We assume all data is the cache per node on solvers of dense matrix operations for exa-scale supercomputers. This is a requirement for actual software in a production run. To reduce communications for a symmetric eigensolver, we executed several communication implementations with either an MPI non-blocking communication or a communication reducing time.

The performance evaluations for up to full nodes of the FX10 system indicate the following: (1) MPI non-blocking implementation was efficient by a factor of 3 in comparison with the baseline implementation, (2) the hybrid MPI execution was a factor of 1.9, (3) our solver was as fast as 2.3x and 22x for a ScaLAPACK routine with optimized blocking size and for cyclic-cyclic distribution, respectively. It is significant that our solver does not use a blocking algorithm for computation, while the current trend of dense solvers, such as ScaLAPACK routines, is to use a blocking algorithm for computation.

Future work includes developing techniques with more communication reducing or hiding the communication for HIT routines, since the ratio of HIT execution to total time increases in execution on full nodes of the FX10. A detailed comparison, both of time and accuracy of the MRRR routine in ScaLAPACK, is also important

---

a By using re-distribution, the distributed matrix can be rearranged with block-cyclic distribution. However, the re-distribution costs may not be marginal, since the matrix is very small in our target. In addition, several parts need to be re-distributed for matrices *A* and *X*. It is known that block-cyclic distribution causes a heavy load imbalance as compared to cyclic-cyclic distribution (*MBSIZE* = 1), if the matrix is very small[5]. Hence, we measure the data with *MBSIZE* = 1 for ScaLAPACK as reference data.

future work. Adaptation of current Auto-tuning technologies[15)-20)] to the proposed communication avoiding method of the eigensolver is also challenging future work in era of exa-scale.

**Acknowledgements**

This work is partially supported by a Grant-in-Aid for Scientific Research (B) "Exa-scale Adaptation to Sparse Iterative Library with Run-time Auto-tuning Facility", No. 24300004, granted by the Ministry of Education, Culture, Sports, Science and Technology (MEXT), Japan. For full node execution, the computational resource of the Fujitsu FX10 was awarded by the "Large-scale HPC Challenge" Project, Information Technology Center, The University of Tokyo.